\documentclass[letter]{ptptex}

\usepackage{graphicx}

\title{
Implication of Classical Black Hole Evaporation Conjecture 
to Floating Black Holes
}

\author{
Takahiro \textsc{Tanaka}
}

\inst{
Yukawa Institute for Theoretical Physics, 
Kyoto University,\\ Kyoto 606-8502,~Japan
}

\abst{
In Randall-Sundrum single-brane (RS-II) model, it was conjectured that there is no 
static large black hole localized on the brane based on adS/CFT correspondence. 
Here we consider the phase diagram of black objects in the models  
extended from the RS-II model. 
We propose a scenario for the phase diagram consistent with 
the classical black hole evaporation conjecture. 
The proposed scenario indicates the existence of a rich 
variety of the families of black objects. 
}

\begin{document}

\maketitle

In the Randall-Sundrum single-brane (RS-II) 
model~\cite{Randall:1999vf}, no stable large black hole 
solution localized on the brane is known. We proposed a conjecture that 
such a large localized black hole solution does not 
exist~\cite{Tanaka:2002rb,Emparan:2002px}, based on adS/CFT
correspondence~\cite{Maldacena,Gubser,Hawking:2000kj}. 
Once gravitational collapse occurs on the brane, the 
collapsed object will form something like a black hole, 
but it should eventually evaporate 
within the classical dynamics. 
In the dual CFT picture 
this evaporation can be interpreted 
as back-reaction due to the Hawking radiation.
Although there might be some possible objections to this 
conjecture~\cite{Fitzpatrick:2006cd}, 
this naive correspondence works pretty well in all known 
examples~\cite{Emparan:2002px,Porrati:2001db,DufLiu,ShiIda,Tanaka:2004ig,Pujolas:2008rc}.  

In the previous studies~\cite{Kudoh:2003xz,Kudoh:2003vg}, 
small black hole solutions localized on the brane have been constructed 
numerically, but the numerical construction of solutions becomes 
more and more difficult as the horizon size increases. 
There are a few analytical works on the existence of small localized 
black holes, but the results are not conclusive 
yet~\cite{Karasik:2003tx,Kodama:2008wf}. 
Our interpretation of this numerical results is that localized 
black hole solutions exist if and only if their size is small 
compared with the bulk curvature scale, $\ell$, 
although recently results which suggest
the possibility that even a small localized 
black hole solution does not exist in a strict sense 
were given~\cite{Yoshino:2008rx}. 


The phase diagram of the black objects in the RS-II model has not been 
clarified yet, but the diagram in the usual KK
compactification (un-warped two-brane model) has been 
established~\cite{Sorkin:2003ka,Kol:2004ww,Kudoh:2004hs}. 
These two models are continuously connected with each other in the 
space of model-specifying parameters, the bulk curvature length $\ell$ 
and the brane separation $d$. 
As we vary these parameters, the phase 
diagram of the black objects will also change continuously. 
Then, there must be a consistent scenario for the phase diagram 
in which only small black hole solutions are allowed in the RS-II model. 

In this letter we discuss the phase diagram of black objects, 
considering models continuously connected to the RS-II model.  
Our basic assumption is that, when the model-specifying parameters are 
continuously varied, a sequence of solutions should 
also change continuously. 
We propose a consistent scenario of the phase diagram, assuming that 
the classical black hole evaporation conjecture is correct. 
\vspace{2mm}

\begin{figure}[t]
\begin{center}
\scalebox{0.67} {\includegraphics{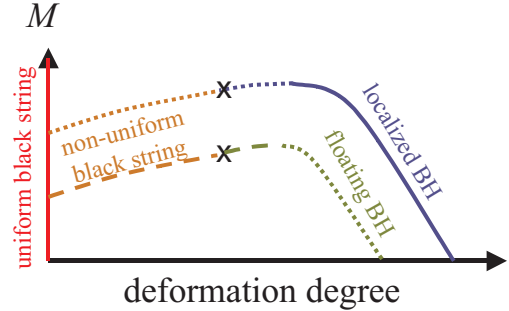}}
\end{center}
\caption{Phase diagram of black objects in the un-warped two-brane
 model. Dashed curves represent unstable sequences.}
\label{fig1}
\end{figure}

\noindent{\it Phase diagram in two-brane model }:
Here we begin with discussing the phase diagram of black objects 
in two-brane model with un-warped bulk. 
The phase diagram is summarized in 
Fig.~\ref{fig1}\cite{Sorkin:2003ka,Kol:2004ww,Kudoh:2004hs}. 
The horizontal axis represents the degree of deformation 
from the uniform black string. The non-uniform black string 
branch starts at the Gregory-Laframme instability point~\cite{GreLaf,Gregory}, 
where the horizon radius measured on the brane is 
as large as the brane separation $d$. 
At some point, 
the sequence of non-uniform black string solutions continues 
to that of localized black hole solutions. 
We refer to this branch as the primary branch. 
In the same plot, we have also shown another curve 
corresponding to the sequence starting at the 
second Gregory-Laframme instability point where the gradient 
of the zero mode has one node in the bulk. 
We refer to this branch as the secondary branch. 
There are infinitely many branches in a similar manner, but 
here we focus only on the primary and the secondary branches. 
In the un-warped case, when we trace a sequence of solutions, 
a non-uniform black string in the secondary
branch is detached from both branes simultaneously. 
On the black hole branch, therefore the black hole is 
floating in the bulk. 

The first question is 
how this diagram is modified once we introduce 
the warp in the bulk by adding a bulk negative 
cosmological constant and appropriate tensions on the branes. 
Notice that 
the primary branch has two solutions,   
depending on which brane has the larger rescaled cross-section 
with the horizon, where the rescaled cross-section means the 
cross-section divided by the squared background warp factor 
evaluated on the brane.
When there is no warp, these two 
solutions are degenerate because two boundary branes are equivalent. 
Once the extra-dimension is warped, however, 
these two solutions are not equivalent any more. 
Here we focus on the solution whose 
rescaled horizon section is larger on the UV brane. 

We point out that the floating black hole 
in the un-warped case cannot stay apart from 
the UV brane when $\ell$ gets smaller. 
This can be understood as follows. 
Here we assume the Randall-Sundrum condition, i.e. 
the brane tension is fine-tuned to accept Minkowski brane. 
The gravity between the 
UV brane, which has positive tension, 
 and a particle floating in the bulk is repulsive. 
To see this, 
we evaluate the acceleration of a test particle fixed at 
a spatial point in the Poincare coordinates, in which 
adS metric is written as 
\begin{eqnarray*}
ds^2=dy^2+e^{-2y/\ell}\left(-dt^2+d{\bf x}^2\right).
\end{eqnarray*}
The acceleration is calculated as 
\begin{eqnarray*} 
 a={(\log g_{tt})_{,y}
   \over \sqrt{g_{yy}}}=-{1\over \ell},
\end{eqnarray*}
which means that the acceleration 
toward the UV brane needed to keep the 
position of a particle fixed in the bulk is 
$y-$independent. 
To have a static configuration, 
the only way to compensate this repulsive (attractive) force
from the UV (IR) brane is the 
self-gravity caused by the mirror images of the particle itself 
on the other 
side of the branes. 

{}From the above observation, we expect that the equilibrium position of
the 
floating black hole should 
move toward the UV brane as $\ell$ decreases. 
However, the attractive force between two black holes 
will be, at most, of $O(1/R)$ with $R$ being the size of the black holes. 
If $R\gg\ell$, the self-gravity 
of black holes will not be sufficient to compensate the repulsive force 
from the UV brane. Thus, a black hole with $R\gg \ell$ cannot float in the bulk. 
In such cases, a black hole on the secondary branch 
necessarily touches the UV brane. 

Now we are ready to discuss the phase diagram of black objects in 
the two-brane model with a warped extra dimension. 
If the topology of the phase diagram is preserved,  
there must be, at least, two 
black hole solutions localized on the UV brane 
for a large horizon radius $(\gtrsim \ell)$, under 
the assumption that the brane separation 
is sufficiently large ($d\gg \ell$). 
We think that the co-existence of two branches of 
localized black holes looks quite unlikely.

\begin{figure}[t]
\begin{center}
\scalebox{0.67} {\includegraphics{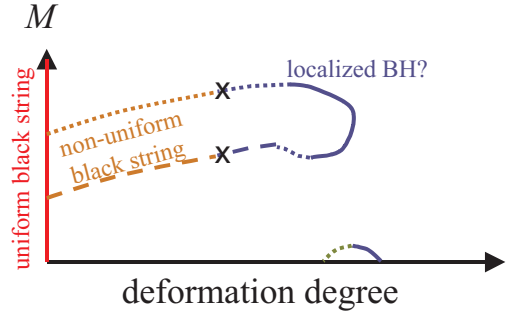}}
\end{center}
\caption{Phase diagram of black objects for two-brane model with warped bulk.}
\label{fig2}
\end{figure}

{}Our basic assumption is that the 
classical black hole evaporation conjecture is correct.  
Then, most likely the phase diagram should be modified 
as shown in Fig.~\ref{fig2}. 
The absence of a large black hole is explained by the 
reconnection between the primary and the secondary branches. 
If it were not for the secondary branch, 
the absence of a large localized black hole 
in the warped case would 
lead to sudden disappearance of a sequence of solutions,  
which would not be understood naturally. 

When the brane separation is small compared with the bulk curvature 
length ($d\ll \ell$), the effect of the warp will not be significant. 
Hence, the phase diagram 
should be topologically identical to that given in Fig.~\ref{fig1}. 
As the parameter $\ell$ decreases, the two branches 
will get closer. At a critical value of 
$\ell(\approx d)$, 
the two branches will touch and interchange.  
As a result, the phase diagram becomes topologically 
as given in Fig.~\ref{fig2}.
\vspace{2mm}

\noindent
{\it Extension to the detuned brane tension }:
Our current discussion can be extended 
to more general cases by considering de-tuned brane tension, 
which is called the Karch-Randall model~\cite{Karch:2000ct}. 
First we consider the case that the deviation from the Randall-Sundrum 
condition is small. We introduce a parameter $\delta\sigma\equiv (6/\kappa_5 \ell)-\sigma>0$, 
where $\sigma$ is the tension of the UV brane. Here we consider 
the large separation limit ($d\to\infty$) for simplicity. 
In order to describe the unperturbed solution with a detuned brane, 
it is convenient to use the coordinates 
\begin{eqnarray*} 
 ds^2=dy^2+\ell^2\cosh^2(y/\ell) 
ds^2_{adS_4}, 
\end{eqnarray*}
where $ds^2_{adS_4}$ is the metric of four-dimensional anti-de Sitter (adS) space 
with unit curvature. 
The brane is on a $y=$constant 
surface, and the value of $y$ on the brane, $y_b$, 
is determined by the condition 
\begin{eqnarray*}
\kappa_5\sigma
=-{6\over \ell}\tanh{y_b\over \ell}. 
\end{eqnarray*}
The limit corresponding to the RS-II model is obtained by setting 
$\sigma=6/\ell$ ($y_b\to -\infty$).
When $\delta\sigma$ is small, we have 
$
 \delta\sigma\approx ({12/ \kappa_5\ell}) e^{2y_b/\ell}.
$
The very outstanding feature for $\delta\sigma\ne 0$
is that the warp factor $\ell^2\cosh^2(y/\ell)$ is not 
monotonic but has a minimum at $y=0$.  
When $\delta\sigma$ is sufficiently small, 
$-y_b$ is very large. Hence, significant deviation from the 
exact RS limit 
arises only in the region distant from the UV brane ($y\gtrsim 0$).  

\begin{figure}[t]
\begin{center}
\scalebox{0.67} {\includegraphics{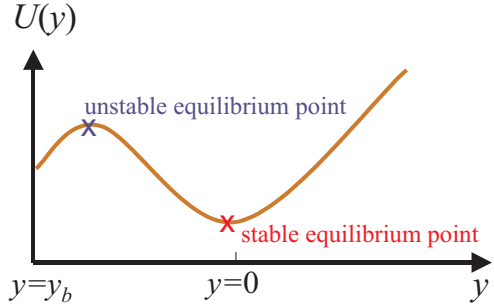}}
\end{center}
\caption{Effective potential for a small particle in the bulk 
in detuned-tension model with adS brane.}
\label{fig3}
\end{figure}

Let us consider a small black hole floating in the bulk. 
Since the acceleration of a static test particle is 
given by $a=(\log\sqrt{g_{00}})_{,y}$, the effective gravitational 
potential (without self-gravity) becomes
\begin{eqnarray*}
 U_{eff}=\log (g_{00}) = \log \left(\ell\cosh {y\over \ell} \right).
\end{eqnarray*} 
The effective potential after taking into account 
the self-gravity will be modified as shown in 
Fig.~\ref{fig3}. 
From this plot, we expect that 
there are two floating black hole solutions when 
the size is small. 
The one close to the UV brane is unstable, 
while the other close to $y=0$ is stable. 
When $\delta\sigma$ 
is small, the stable floating black hole is 
very far from the UV brane. 
In the limit $\delta\sigma\to 0$, this black hole 
is infinitely far. As a result, 
this sequence of solutions disappears from the phase 
diagram of the RS-II model.  
The distance from the UV brane to 
the unstable equilibrium point will not be sensitive 
to the small change of $\sigma$.  
Hence, this branch is smoothly connected to the diagram shown in 
Fig.~\ref{fig2}. 

In order to draw the phase diagram in the regime 
$0<\delta\sigma \ll 1/\kappa_5\ell$, it will be important 
to know how the branch of stable floating black holes 
extends to a larger size. It will be easy to imagine 
that this sequence also touches the UV brane when 
the area of the five dimensional horizon becomes 
sufficiently large.  
Again adS/CFT correspondence provides us with 
a method for estimating 
the critical size at which the floating black hole touches the UV
brane. 

In the asymptotically flat case 
we cannot construct a static quantum  
black hole solution due to the quantum 
back-reaction~\cite{Fabbri:2005zn}. 
In this case the back-reaction 
is too strong to keep the asymptotic spacetime structure unchanged. 
However, it is not the case in asymptotically adS spacetime~\cite{Hawking,Galfard:2005va}. 
One possible choice of static quantum state in which 
the energy density on the event horizon is regular is   
the thermal Hartle-Hawking (HH) state. 
In the asymptotically flat case, the HH 
state leads to a constant energy density in the asymptotic 
region. Hence, the total mass diverges. 
In contrast, in the case of asymptotically 
adS spacetime 
vacuum energy density exists from the beginning. As a result,  
the lapse function 
does not converge to a constant value at a large $r$. 
In fact, the lapse function of the background four-dimensional adS space 
is given by $\sqrt{f}$ with
\begin{eqnarray*}
 f=1-{(2\kappa_4 M/ r)}+{(r^2/ L^2)}, 
\end{eqnarray*}
where $L(\ll \ell)$ is the four-dimensional adS curvature scale. 
Since the temperature red-shifts in proportion to $1/\sqrt{f}$,   
the energy density
decreases very rapidly for $r\gg L$. 
When the size of the black hole is large, the black hole temperature 
is low. Therefore the energy density of CFT 
becomes important only at a large distance. 
If this scale is larger than the adS curvature scale $L$, 
the back-reaction effect is cut off due to the red-shift factor. 
Therefore there is a possibility of having a static large black hole 
configuration consistent with the back-reaction due to CFT. 
To the contrary, if the size of the black hole is not large, 
the back-reaction becomes important below the adS curvature scale. 
Then, a static quantum black hole solution will not exist
as in the asymptotically flat case. 
This means that there is a critical size 
beyond which a static quantum black hole solution exists. 

\begin{figure}[t]
\begin{center}
\scalebox{0.67} {\includegraphics{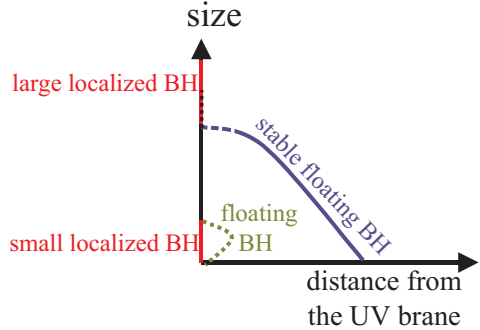}}
\end{center}
\caption{Phase diagram of black objects 
in detuned-tension model 
with adS brane.}
\label{fig4}
\end{figure}

Following the above picture, we can estimate the minimum 
size of a large static black hole in adS space. 
We quote the results from Refs.~\citen{Hawking}. Substituting 
the effective number of species 
$\approx \ell^2/\kappa_4$
that adS/CFT correspondence tells, 
we find that the back reaction becomes important when 
$M <\sqrt{\ell L}/\kappa_4$. 
This result suggests that the sequence of four dimensional 
large localized black holes with a small value of 
$\delta\sigma$ 
starts with $M\approx\sqrt{\ell L}/\kappa_4 $. 
In the five dimensional picture this means that the 
sequence of stable floating black holes should touch the 
UV brane when the size of the black hole is as large as 
$\sqrt{\ell L}$. 
Thus, the phase diagram in this regime will 
be like Fig.~\ref{fig4}. (Here black string configuration
as discussed in Ref.~\citen{Gregory:2008br} is not taken into account.)

There must be a sequence of solutions on the four dimensional CFT side 
which corresponds to the sequence of stable floating 
BHs that is expected to
exist on the 
five dimensional bulk gravity side. 
The four dimensional counterpart should be 
a sequence of star solutions composed of CFT matter. 
Such CFT stars will be well approximated by stars made of 
radiation fluid, 
which was already studied long time ago 
by Page and Phillips~\cite{Page:1985em}. 
If we substitute the effective number of species 
$\approx \ell^2/\kappa_4$ into their results, we find that 
the sequence of star solutions terminates where the mass of star is 
$O(\sqrt{\ell L/\kappa_4})$. The end point of the sequence 
is a singular solution which has diverging central energy density. 
Diverging density means diverging temperature, 
which indicates that the lapse function also vanishes at the center 
in this limiting case. 
The corresponding five dimensional 
metric should also have vanishing lapse function at the center on 
the brane. For a static black hole spacetime, the lapse 
function will vanish only on the event horizon. Hence, we can 
imagine that the moment when the bulk floating black hole just 
touches the brane in the five dimensional bulk gravity picture 
corresponds to the end point of the sequence of CFT stars 
in the four dimensional CFT picture. At this moment the size of 
the five dimensional black hole is estimated from the entropy 
of the corresponding CFT star to be $O((\ell L)^{3/2})$. 
How the sequence continues after 
formation of black hole in the four dimensional CFT picture 
will be discussed in the
forthcoming paper~\cite{Kashiyama}.  

Let us further reduce the brane tension. 
When $\sigma\approx 0$, $y_b$ is close to $0$. In the 
most of region in the bulk, the repulsive force from the 
UV brane is screened by the attractive nature of the 
bulk negative cosmological constant. As a result, 
a test particle feels net 
repulsive force from the brane only in the limited 
small region near the brane. Thus, only a small black hole 
which can fit within this tiny region can float in 
the bulk. 

In the limit $\sigma\to 0$ the floating branches  
are not allowed at all. The size of the horizon measured 
on the brane at the transition point becomes zero 
in this limit. As a result, one sequence of black hole solutions 
localized on the brane remains. In this limit, this sequence of 
solution is nothing but adS-Schwarzschild solution cut 
by a tensionless brane placed on the equatorial plane. 
It is already proven that 
there is no branching point (= solution with a zero mode) 
along this sequence~\cite{Kodama:2004kz}. This fact is completely 
in harmony with our phase diagram. 

To summarize, we have shown it possible to describe 
a scenario of the  
phase diagram evolution of black objects in models connected to RS-II 
brane world model, with a small number of assumptions.  
The obtained phase diagram is perfectly 
consistent with adS/CFT correspondence.  
An interesting point to stress is that 
in the Karch-Randall detuned tension brane world 
the existence of large static black holes localized on the brane
is naturally expected in contrast to RS-II model.
We will apply the numerical methods used to investigate black hole 
solutions in the RS-II model~\cite{Kudoh:2003xz,tanahashi} 
to the Karch-Randall case in the future work. 
\vspace{2mm}

The author thanks R. Emparan, A. Flachi , N. Kaloper
and N. Tanahashi for valuable conversation. 
This work is partially supported
by Grant-in-Aid for Scientific Research, Nos. 19540285 and 17340075. 
The author also acknowledges the support from the Grant-in-Aid for the Global COE Program
``The Next Generation of Physics, Spun from Universality 
and Emergence'' from the Ministry of Education, Culture,
Sports, Science and Technology (MEXT) of Japan.

\end{document}